\documentstyle[aas2pp4]{article}

\lefthead{Alcock et al.}
\righthead{Difference Image Analysis}

\begin{document}
\title{Difference Image Analysis of Galactic Microlensing}

\begin{center}
{\large\bf I. Data Analysis}
\end{center}

\author{C. Alcock\altaffilmark{1,2}, R.A. Allsman\altaffilmark{3}, D. Alves\altaffilmark{1,4},
T.S. Axelrod\altaffilmark{5}, A.C. Becker\altaffilmark{6}, D.P. Bennett\altaffilmark{1,2},\\
K.H. Cook\altaffilmark{1,2}, A. J. Drake\altaffilmark{5}, K.C. Freeman\altaffilmark{5}, K. Griest\altaffilmark{2,7},
M.J. Lehner\altaffilmark{8}, S.L. Marshall\altaffilmark{1,2},\\ D. Minniti\altaffilmark{1,13}, 
B.A. Peterson\altaffilmark{5}, M.R. Pratt\altaffilmark{9}, P.J. Quinn\altaffilmark{10},
C.W. Stubbs\altaffilmark{2,5,6},\\ W. Sutherland\altaffilmark{11},
A. Tomaney\altaffilmark{6}, T. Vandedei\altaffilmark{7}, and D.L. Welch\altaffilmark{12}}

\author{\bf(The MACHO Collaboration)}

\altaffiltext{1}{Lawrence Livermore National Laboratory, Livermore, CA 94550} 
\altaffiltext{2}{Center for Particle Astrophysics, University of California, Berkley, CA 94720} 
\altaffiltext{3}{Supercomputing Facility, Australian National University, Canberra, ACT 0200, Australia}
\altaffiltext{4}{Department of Physics, University of California, Berkeley, CA 95616}
\altaffiltext{5}{Mount Stromlo and Siding Spring Observatories, Weston Creek, Canberra, ACT 2611, Australia}
\altaffiltext{6}{Department of Astronomy and Physics, University of Washington, Seattle, WA 98195}
\altaffiltext{7}{Department of Physics, University of California, San Diego, CA 92093}
\altaffiltext{8}{Department of Physics, University of Sheffield, Sheffield s3 7RH, UK}
\altaffiltext{9}{Center for Space Research, MIT, Cambridge, MA 02139}
\altaffiltext{10}{European Southern Observatory, Karl Schwarzchild Str.\ 2, D-85748 G\"{a}rching bel M\"{u}nchen, Germany}
\altaffiltext{11}{Department of Physics, University of Oxford, Oxford OX1 3RH, UK}
\altaffiltext{12}{Department of Physics and Astronomy, McMaster University, Hamilton, ON L8S 4M1, Canada}
\altaffiltext{13}{Departmento de Astronomia, P. Universidad Cat\'olica, Casilla 104, Santiago 22, Chile}

\begin{abstract}
  
  This is a preliminary report on the application of Difference Image 
  Analysis (DIA) to galactic bulge images. The aim of this analysis 
  is to increase the sensitivity to the detection of gravitational 
  microlensing.

  We discuss how the DIA technique simplifies the process of discovering
  microlensing events by detecting only objects which have variable flux. We
  illustrate how the DIA technique is not limited to detection of so called
  ``pixel lensing'' events, but can also be used to improve photometry for
  classical microlensing events by removing the effects of blending. We will
  present a method whereby DIA can be used to reveal the true unblended
  colours, positions and light curves of microlensing events.

  We discuss the need for a technique to obtain the accurate
  microlensing time scales from blended sources, and present a possible 
  solution to this problem using the existing HST colour magnitude 
  diagrams of the galactic bulge and LMC. The use of such a solution with
  both classical and pixel microlensing searches is discussed.

  We show that one of the major causes of systematic noise in 
  DIA is differential refraction. A technique for removing 
  this systematic by effectively registering images to a common airmass 
  is presented. Improvements to commonly used image differencing 
  techniques are discussed.

\end{abstract}


\section{INTRODUCTION}

The study of microlensing has become well established after a number of
groups followed up the ground breaking proposal of Paczy\'nski (1986).
Over the past few
years well over a hundred events which can only reasonably be attributed to
microlensing have been discovered by the MACHO (\cite{alc98}), EROS
(\cite{bea95}), DUO (\cite{ala97c}) and OGLE (\cite{pac94}) groups, in their
attempt to characterise the nature of the dark matter halo of our galaxy.

Fits to the light curves from microlensing events do 
not give us a direct measure of the mass of the objects causing the
magnification. What they do provide is a measure of the
lensing time scale and amplification. The amplification of light from 
a point source due to microlensing is given by equation \ref{ampl}, where
$u(t)$ can be obtained from equation \ref{impact}, 
$u_{min}$ is the impact parameter in terms of Einstein radii 
and $t_{max}$ is the time at which maximum amplification is reached.

\begin{equation}
A(t)= \frac{u^{2}+2}{u(u^{2}+4)^{1/2}}
\label{ampl}
\end{equation}

\begin{equation}
u^{2}(t) = u^{2}_{min} + \biggl(\frac{2(t- t_{max})}{\hat t}\biggr)^{2}
\label{impact}
\end{equation}

The time scale $\hat t$ of a microlensing event is set by 
the time the Einstein ring takes to traverse the source star
at velocity $V_{\perp}$. The projected size of the Einstein radius
$R_{E}$ (equation \ref{chiswolson}), is dependent upon $D_{d}$,
the observer-lens distance, $D_{s}$, the source-observer 
distance, $D_{ds}$, the lens source distance, and the lens mass 
$M$ (in solar masses).

\begin{equation}
R_{E} = \sqrt{\frac{4GM}{c^{2}}\frac{D_{d}D_{ds}}{D_{s}}}
\label{chiswolson}
\end{equation}

To uniquely determine the mass of the lens causing any given event, one must
determine its distance, the source distance and its transverse velocities
relative to our line-of-sight. This is usually not possible.
However, for a small number of exotic events where binarity or parallax are 
found, the degeneracy of $M$, $D_{d}/D_{s}$ and $V{\perp}$ (e.g. \cite{alc95})
can be broken. One can instead
extract the distribution of lens masses by assuming the distribution of 
lens transverse velocities and distances (\cite{gri91}). As this is a 
statistical process, an increase in the number of events leads to 
increased accuracy in the determination of lens mass distribution.


The number of events detected in microlensing surveys is limited by the
number of stars which it is possible to monitor and the overall detection
efficiency. In terms of telescopes, this means such surveys are limited by
the size of the field-of-view of the survey telescope and its light
gathering power. Limitations to such surveys also come from sampling rate
and the seeing. These effects limit the accuracy of the photometry obtained
and thus the event detection threshold. To maximise the number of events
one naturally chooses to observe fields with the greatest density of
stars. This introduces crowding which sets a seeing dependent limiting
magnitude for such surveys. To maximise the number of monitored stars one
can take an image in the best seeing conditions, where the crowding and sky
background levels are at minimum. The detected stars can then be monitored
even when the seeing is poor and sky background is high, by transforming the
coordinates of stars from the reference image to each subsequent
observation. This is the standard approach taken in such surveys.

The use of crowded fields limits the accuracy of photometry. For
images with poor seeing conditions, faint stars become immersed in 
the flux from neighbouring brighter stars. An important consideration 
when working with such fields, is that only a small fraction of the 
stars are actually detected. In fact, within each seeing disk there 
are generally many stars. The determined stellar centroid is actually 
a flux weighted mean centroid of the seeing
disk. Any one of the stars within such a blend can be gravitationally lensed.
Blending thus has a major effect on the actual number of stars that are
monitored and hence the optical depth to microlensing.
Such blending of microlensing events causes an amplification 
bias in the number of detected events (\cite{bal93,nem94,han97a,woz97,ala97a}).
In past analyses such effects have been taken into account statistically (\cite{alc97}),
but corrections are difficult to apply on an event by event basis
(\cite{han97a}). In this way the method of monitoring a star field 
using fixed positions, in some sense, does allow us to monitor some 
sources which were too faint to have been detected by virtue of their
own flux. This does not, however, allow us to detect any microlensing 
events due to such source stars which are not blended with a 
neighbouring star or stars bright enough to be detected.

New, large, ground based telescopes are useful in increasing the
number and frequency with which star fields can be monitored because of 
shorter exposure times are required. But such telescopes
are still prone to the same intrinsic seeing limits as small telescopes, 
and thus the same crowding limits of images. The effective seeing can be 
improved with the use of better observing sites, but 
this does not resolve the issue of blending completely when many stars 
are present per square arc second. Attention must be applied to improving the 
reduction techniques to remove blending effects. Such techniques 
would also ideally increase the number of events detected by increasing 
the effective number of stars monitored per image. To achieve both requires 
a technique which will overcome the crowding in the images and 
increase the monitored area within each image. With this intent we shall
apply difference image analysis (hereafter DIA\footnote{This technique 
is also referred to as image subtraction.}). This uses a similar technique 
to that first described by Crotts (1992) and Phillips \& Davis (1995).

Lensing events where the source stars are only detectable during
microlensing are usually designated `pixel lensing' events.
Events where the source star is a resolved are termed  `classical
microlensing' events (see \cite{gou96} and \cite{gou97} for full
details). The term pixel lensing comes from an analogy with 
microlensing surveys in M31 (\cite{ans97}), where each pixel in 
this survey represents the sum of hundreds of unresolved stars that 
may be microlensed. This term is applicable with local 
microlensing surveys (bulge, LMC) to an extent, since even for the bulge, given 
arc second seeing, each pixel contains light from a number of stars. 
In this way we define local pixel lensing
events as microlensing events where the sources are not associated 
with stars we monitor. Furthermore, to remove the ambiguity in 
what is a pixel lensing event and what is a classical lensing event,
we will use the colour and position information available with 
the DIA technique. The three properties {\em flux, colour, position} 
allow us to make a fairly robust separation between these event types.

In the following sections we will demonstrate how the DIA technique can be
used to increase the number and quality of results from microlensing
surveys. In the next section we will briefly outline the observational
strategy. In \S 3 we will also discuss how the DIA technique was employed
on our set of data and what improvements to the standard method were made. 
In \S 4 we will discuss the relevance of DIA to the issue of 
blending. In \S 5 a comparison between PSF photometry and this 
technique is made. In \S 6 we outline how to determine source fluxes
from the results. In the final section we shall make our 
concluding remarks.
{\it The light curves and parameters for the microlensing 
events discovered will be presented in} \cite{alc99}.

\section{OBSERVATIONS}

The MACHO observation database consists of over 70,000 individual
observations of the galactic bulge, the LMC and the SMC. The present
analysis considers only a single $42^{\prime}$ by $42^{\prime}$ field (Macho
id 108) in the galactic bulge centred at $\alpha= 18\arcdeg 01'20''$,
$\delta = -28\arcdeg17'39''$ $(J2000)$. Observations were taken on the Mount
Stromlo and Siding Spring Observatories' 1.3m Great Melbourne Telescope
with the dual-colour (red, blue) wide-field ``Macho camera''. All bulge observations
have $150$ second exposure times.

The Macho camera consists of a mosaic of eight 2k $\times$ 2k CCDs (four red, 
four blue) with $0.63''$ pixels. Each of the eight CCDs in the Macho 
camera consists of two amps.
Light passing through the telescope is 
separated into two passbands using a dichroic beam splitter allowing both
red and blue images of a field to be taken simultaneously (\cite{mar93}).

Observations in this analysis span the dates from the 10th of March 1995 to
the 2nd of August 1997 with breaks from the end of October to the beginning of
March each year when the bulge is unobservable at Mount Stromlo.
Within this observing period $385$ observations of the target field were taken. 
The seeing for the data set varies from $1.2^{\prime\prime}$
to $6.5^{\prime\prime}$ with a mean of $2.3^{\prime\prime}$. 
The sky background level in the blue band-pass varies from $1000$ to $30000$ 
counts with a mean $\sim2600$ counts. Similar levels were observed for red 
images.

We chose to reject a number of observations from this dataset as
good photometry was required to detect of microlensing of faint stars.
Images with seeing FWHM $> 4^{\prime\prime}$ or blue-band sky background $> 8000$ ADU were 
excluded from the reduction. These two cuts removed 42 observations from 
the data set.
As the usable area of a difference images is dependent on the 
differences in pointing between observations, 
we also excluded a small number of images where the difference in 
pointing of the reference and subsequent 
observation was greater than $25^{\prime\prime}$.
With this criteria imposed we rejected 19 more observations leaving
324 before reduction was attempted. 

\section{DIFFERENCE IMAGE ANALYSIS}\label{dia}

Since the work of Crotts (1992) and Phillips \& Davis (1995) there have been
a number of applications of DIA type techniques (\cite{tom96,rei98,ala97b}).
These papers show the rapid evolution of the technique to a near optimal
case. The implementation of the technique as used here varies in several
important aspects to these approaches. So we shall outline our technique,
noting the similarities and differences to these applications.

In brief, DIA of our images involves registering a test image to a
preselected, high signal-to-noise ratio (hereafter S/N), low airmass,
stacked reference image. Next $\sim\! 200$ bright, uncrowded, so called PSF
stars, are selected to calculate the convolution required to map the
template image to the test one. The convolution kernel for this mapping is
calculated and applied to the reference. The test image is then
photometrically normalised (scaled and offset) to match the convolved
reference image. The resultant, registered, convolved and normalised image
is then subtracted from the test image. The process is carried out for red
and blue passbands, and objects are detected.  The positions of these
objects are matched in the two colours to separate real objects from spurious
ones. The results are then sorted and characterised to separate 
microlensing events from variable stars.

\subsection{Template construction}

To perform DIA the first requirement is to have a comparison {\em reference
image} (template) to difference against. To minimise the noise
contribution to the final difference images from such a reference image, it
is an obvious step to use the highest S/N image. The MACHO project has a
database of hundreds of observations of each field taken with the same
telescope, under similar observing conditions. This database makes it
possible to increase the S/N for a reference image simply by stacking
matched images. As the the highest S/N is associated
with the best seeing, we choose to degrade the reference image seeing 
to the test image. Since we require that the reference image has good
seeing we can only combine images with good seeing to create it.
This considerably limits the number of images we can
combine to make the reference image. 

To produce this reference image we require the constituent images be taken at 
low airmass. This is required because the combination must be unaffected
by refraction (see section~\ref{diffrac}). We also require that such images do not have high
sky backgrounds or sky gradients, so that the combined sky level contributes
little noise to the final difference images. Low sky levels allow us to
place the lowest possible detection thresholds for pixel lensing events.
Care has to be taken selecting images, as even a small residual sky 
gradient in the combined reference image can affect how well two images are
photometrically normalised and thus the quality of the entire set of
difference images produced.

Aside from the advantage of an increased S/N in the combined reference
image, the process of stacking images has the effect of removing most of the
bad pixel regions in the images. Small features in the images
due to differences in sky background, transparency, fringing, or poor
flatfielding are also reduced.

The airmass, seeing, and sky level selection criteria strongly restrict the 
number of observations suitable for the reference image combination. 
Among our $\sim324$ observations, only five meet all our requirements.

\subsection{Image Registration}\label{allign}

The importance of obtaining an accurate registration when performing DIA can
not be over emphasised. To accomplish an accurate registration, and to
reduce the time required for analysis, we took initial estimates for the
geometric alignment transformations from a database produced when each image
was PSF photometered. These transformations were applied to some hundreds of
bright stars spread across each image. The transformed positions of these
stars were then used as starting points where more accurate centroid
positions were determined. The initial centroid positions were found to be
good to within a pixel ($0.63''$). Such accuracy means we do not have little
possibility of centering on the wrong star. These new centroid positions for
the stars in the test and reference images were used in the IRAF tasks {\em
geomap} and {\em geotran} to register the image. The use of a large number
of stars allowed us to determine an accurate mapping between the images
where small rotations, translations and distortions may be present.

The images were initial registered with a simple linear transformation on a
$10'\times 10'$ scale. However, we found this was inadequate for the alignment
accuracy we require from this technique ($< 0.1^{\prime\prime}$). This is
likely to be caused by distortions between images, due to small effects such as
differences in telescope flexure with zenith distance.  For the
final analysis we have used a fifth order Chebyshev geometric transformation 
mapping for blue images, and a third order mapping for red images (where 
differential refraction effects are not so important, see next section).

The pointing accuracy of the Great Melbourne Telescope is good, with the
standard deviation in offset between successive observations of a field
being approximately $10''$.  For difference images this results in the loss
of a very small amount of data at the edges of the images where the
observation and reference image do not overlap.

\begin{figure}
\epsscale{1.0}
\plotone{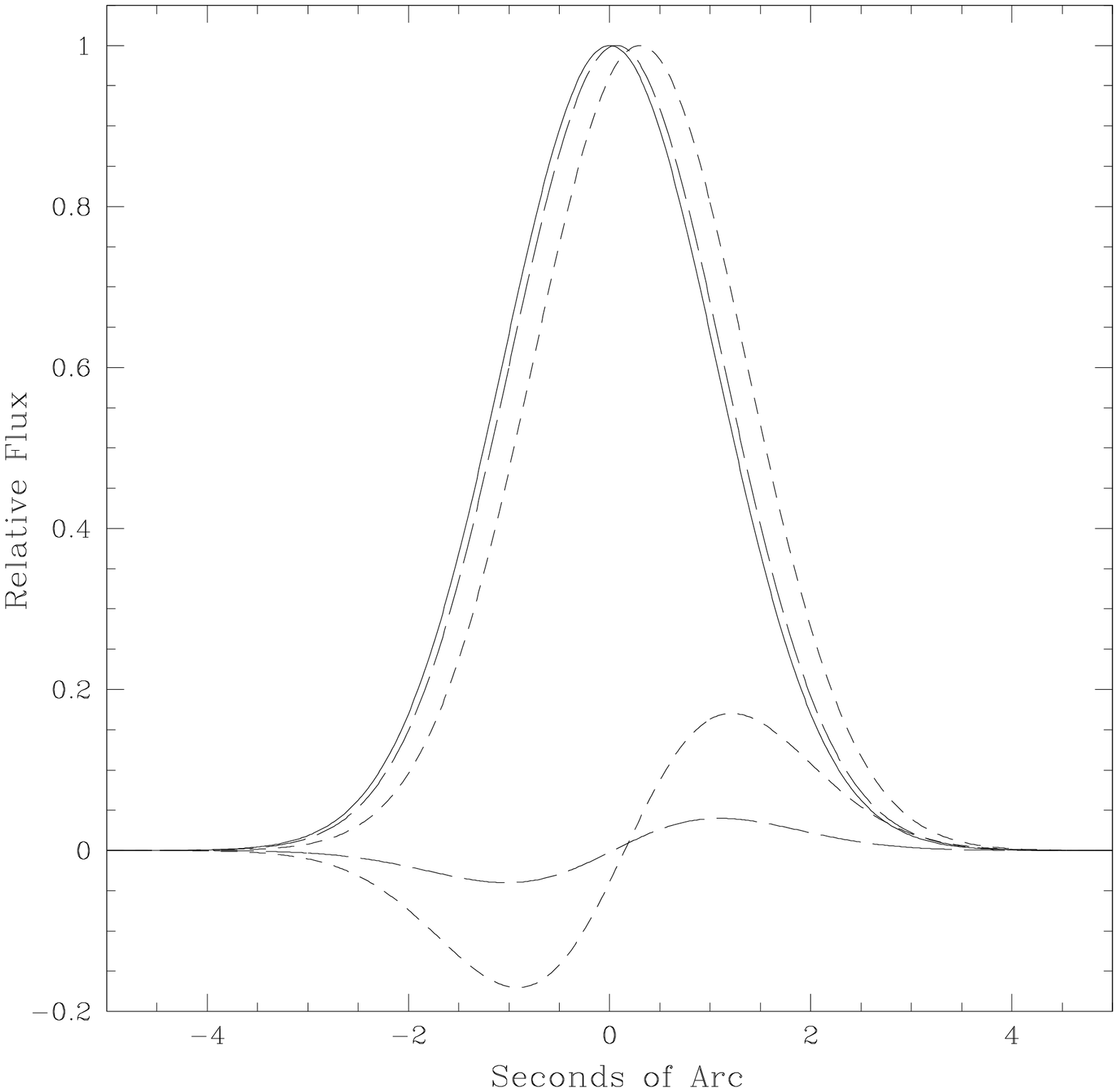}
\caption{Gaussian modelled profiles of three stars and the 
residuals when the two dashed profiles are differenced against 
the solid one. The short dashed lines corresponds to a centroid 
offset of $0.3^{\prime\prime}$. The long dashed lines corresponds 
to an offset of $0.07^{\prime\prime}$, the average offset of our data.
\label{fig1}}
\end{figure}

The registration process is accurate to about 0.07 pixels ($\sim 45$ milli-arc seconds) on
average (c.f.  \cite{rei98}, $\sim0.3''$).  We believe our accuracy is due
to the inclusion of higher order transformation terms and the large number
of stars used to constrain the fit. To show the importance of accurate image
registration we present figure \ref{fig1}. In this figure we approximate a
standard stellar profile as a Gaussian. Using the average seeing of
$\sim2.5^{\prime\prime}$ and an offset between two stellar profiles of
$0.3''$, we expect the average residual (as seen perpendicular to the offset
direction) to have the form of the long dashed line given. For our average
registration accuracy ($0.07''$) the stellar and residual profiles are given
by the short dashed lines. For this figure we did not include the effects of
pixelisation or photon noise, as this makes little difference to the
observed effect.

\placefigure{fig1}

\subsection{Differential Refraction Corrections}\label{diffrac}

The first mention of the deleterious effects of atmospheric refraction on
difference images was made by Tomaney and Crotts (1996). They stressed that,
for broad-band filters, the centroid position of a star is dependent upon
the airmass of the observation. The phenomenon is well known in astrometry
and has the effect of increasingly elongating the PSF with airmass.
For observations taken at different airmasses this leads to poor subtraction. By poor
subtraction we mean that the systematic noise contribution from the
refraction effect is obvious in the images.
The presence of this is revealed by a significant residual flux remaining in
the difference image.

To better understand the refraction effect, one must understand that every
star within the frame has a centroid position which is dependent upon its colour and
the airmass at which it is observed (\cite{Gub98}). If all the stars were of
the same colour then the registration process alone would compensate for 
the shift in centroid position with airmass, and the difference image would 
only be effected by the elongation. However, this is not the case. The image registration is in
reality a registration to the average colour of the stars.  We suggest that,
two images observed through broad-band filters at blue wavelengths and at
moderately different airmasses, can not be geometrically registered 
accurately enough to provide truly photon noise limited difference images. 
However, we further suggest that, if each star were to be registered 
separately we could come close to this.
With this realisation in mind, Tomaney and Crotts (1996) proposed that
`There was no easy way to compensate for this, particularly without
knowledge of the precise position and colour of all detected and undetected
(blended) stars'. Similar effects have also been noted by
Melchior et al. (1996) who believed that atmospheric dispersion affected
their pixel lensing search in M31. This effect was given to be between
$0.73''$ to $2.75''$ in the extremes of their blue filter. With this result
it is easy to understand the importance such an effect could have.  However,
such a large effect does not occur in our DIA analysis, even in the worst 
case, as the first order refraction effect is an offset of all stars in
an observation. This offset
is taken out in the registration process. It is only the smaller second
order effects, caused by differences in star colour which cause differential
refraction offsets. The corrections for these effects are in the
order of tenths of an arc second at blue band wavelengths and so 
are still appreciable.

To correct for differential refraction the following approach has been
taken. As a first approximation, the effective temperature for any given
star can be associated with a colour. We know the position of the centroid
of a star is dependent upon the colour of the star and the airmass. So if one
knows the airmass, the colour and the observation parallactic angle it 
should be possible to determine the change to the stellar flux 
distribution with airmass and parallactic angle.

\begin{figure}
\epsscale{1.0}
\plotone{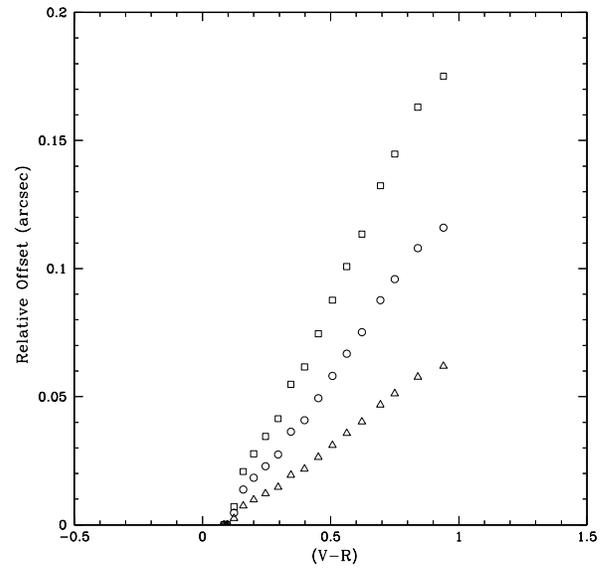}
\caption{\label{fig2} Predicted differential refraction offsets between 
stellar centroids for a range of star colours and airmasses. Offsets 
are simulated for three airmasses relative to a reference airmass of 1.01 
and V-R colour of zero. The three airmass values are 1.9 (squares), 
1.5 (circles) and 1.2 (triangles). Results are for the MACHO Blue ($B_{m}$) 
passband.}
\end{figure}

\placefigure{fig2}

To determine the degree of change we first modelled our galactic bulge
stars based on Bessell's (\cite{Bes98}) tabulated colour-temperature
relations. We then assumed a blackbody spectrum approximation
for these stars with no blanketing or spectral features.

These spectra were then integrated within the MACHO Camera's blue($B_{m}$)
and red($R_{m}$) passband responses to find the average centroid wavelength
associated with a given star colour. The V-R colours for Bessell's (1998)
tabulated stars were transformed to Macho $R_{m}$ and $B_{m}$ colours using
the results of Alves (1998). A star of a given colour can thus be related
to this {\it centroid wavelength}.

\begin{equation}\label{const}
R_{0} = \frac{(n^{2}-1)}{2n^{2}}
\nonumber
\end{equation}

\begin{equation}\label{refr}
R = z_{t} - z_{a}
\end{equation}

\begin{equation}\label{norm}
R \approx R_{0}tanz_{a} - 0''\!.067tan^{3}z_{a}
\nonumber
\end{equation}

The refraction of light at a given
wavelength, temperature and airmass is given by \cite{fil82}. Using 
the equations in \cite{fil82} and equation \ref{const}
one may determine the constant of refraction, where $n$ is the refractive
index of air. The degree of refraction is given by equation\ref{refr}, where $z_{t}$
is the true zenith position and $z_{a}$ the apparent. Under normal
temperature and pressure conditions ($15\arcdeg$C, $760$ mm Hg) this
reduces to equation \ref{norm}, and $R_{0}$ becomes $58.3''$.

The positional shift of the centroid of the star is associated its average
blackbody wavelength. For a star of known colour the centroid shift can be
obtained. The offset in the centroid positions for the model stars are 
given in figure \ref{fig2} for three values of airmass. To make use of this
information, one needs to know the colour of each star and must be able to
shift its centroid in proportion to this colour. This task sounds more
daunting than it really is. In fact, this can easily be achieved (at least
approximately), for two images taken at a known airmass in the following
way. Using the IRAF tasks {\em geomap} and {\em geotran}, it is possible to
map two images of different sizes, orientations and even geometric
distortions onto each other. We used these tasks to map the images from the
two passbands of the MACHO camera onto each other. These mapped images were
then used to form a quotient image which, after sky background subtraction,
can serve as a colour map of each pixel in the field. However, as the
centroid wavelengths of each star within the two passbands varies with
airmass, one must calculate this quotient near unit airmass not to have a
colour map which is affected by refraction.

\begin{figure}
\epsscale{1.0}
\plotone{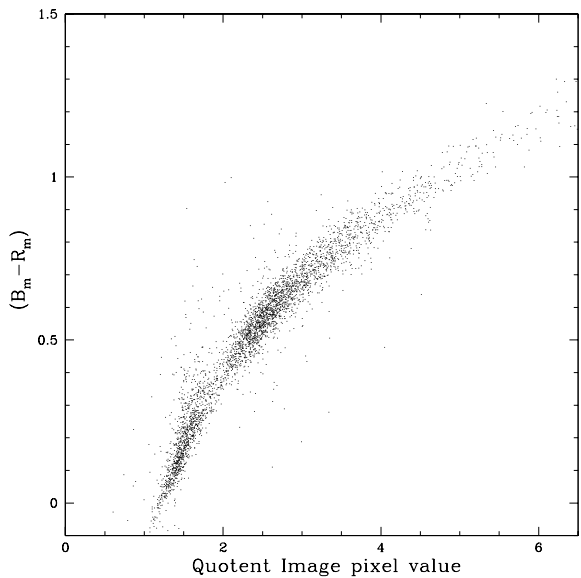}
\caption{The relationship between pixel values in the red divided by blue 
quotient image (colour map) and the standard MACHO passband colours
$B_{m}$ and $R_{m}$. Values are used to determine and correct for
offsets cause by differential refraction for each pixel(star) in the 
image.\label{Col}}
\end{figure}

To find the relationship between the pixel values of our colour map, and the
V-R colours of the stars present in the colour map, a simple calibration is
carried out. Having performed photometry on the individual images in the
colour map, one can associate a $B_{m}-R_{m}$ and hence, a V-R colour, with
a quotient image pixel value. We thus have a V-R colour for every pixel
within the image (except for saturated or bad pixels). The
relationship between the measured $B_{m}-R_{m}$ photometry values and 
the associated colour map pixel value is given in figure \ref{Col}.
The relationship is quite strong for most stars. A small number of 
points are scattered due to stars being blended with neighbours.

\placefigure{Col}
 
\begin{figure}
\epsscale{1.1}
\plotone{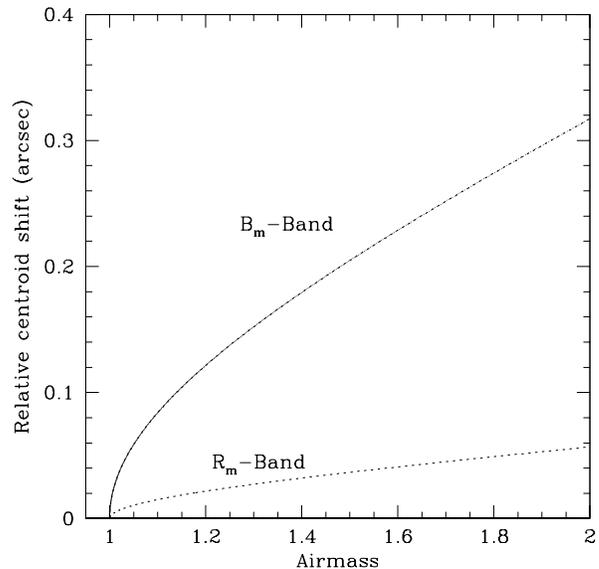}
\caption{Difference in the offset induced by differential 
refraction as a function of airmass, for a 5K (model) star 
and a 10K (model) star. The two MACHO Camera passbands $B_{m}$ and $R_{m}$
are shown. Offsets in the blue passband are much bigger than the red 
because of the strong wavelength dependence of refraction.\label{fig4}}
\end{figure}

As the degree of refraction is wavelength dependent it is in fact only
necessary for us to apply our corrections to images taken through the blue
passband. In red passband the centroid offset is less than one quarter (\(<
0.1''\)) that of the blue passband (see figure \ref{fig4}).

\placefigure{fig4}

The final result is that each blue image pixel value in the reference image
was interpolated based on the calibrated colour map pixel values, the airmass
of the observation relative to the reference image, and the parallactic
angle of the observation. The flux corrections were carried out while
imposing a condition of flux conservation within the image.  We interpolated
the flux in the reference image rather than in each observation as, for most
observation it is not possible to make a colour map for these due to first
order of refraction effects. Colour maps made with individual observations
would have much lower S/N and CCD defects.

The results from this reduction were used as an initial estimate, as it was
known that the results for the differential refraction offsets would be
dependent on the assumptions about the model stars (smooth blackbody
emission) and the uncertainty in the form of the Macho blue band response
function. This estimate was improved by differencing a number of images
taken at a range of airmasses to obtain a semi-empirical result for the
offset required with colour and airmass (relative to the reference image).

\placefigure{figdiffl}
\notetoeditor{This figure should span a page.}

The scale of our corrections can be seen in figure \ref{figdiffl}.  On the
left is a difference image where the effects of differential refraction have
not been removed. On the right the differential refraction corrected images is
shown. The test image used in producing these difference images is at an
airmass of 2.4 (approximately the airmass limit imposed on these
observations). The template airmass is 1.01, so the corrections applied in this
case are quite large. However, the same degree of differential refraction 
effects is observed at much lower airmasses when the seeing quality of 
the test image is good. The residual is highly dependent on seeing, with 
the best seeing images being affected much more because of the steepness 
of the stellar profile.

The calibrated differential refraction corrections for the blue passband of
the Macho camera, as a function of airmass and colour, are given by:

\begin{equation}\label{norm2}
S = 0.29 \times [tan(acos(1/A_{R}) - tan(acos(1/A_{T}))]
\end{equation}

\begin{equation}\label{norm3}
O = S \times (C - 0.6)
\end{equation}

S is related to the scale of the offset for airmasses $A_{R}$ and $A_{T}$,
of the reference and test images respectively. $O$ is the offset to apply
to a given pixel (in arc seconds) of V-R colour $C$. This process compensate
for the shift in the stars centroid position. It does not compensate for the
dispersion of the PSF with airmass. However, we believe this effect to be
quite small and for our purposes ignorable.

\subsection{PSF Matching}

The Point Spread Function (PSF) shape of two images taken at different times
is never exactly the same. If we were to match the photometric conditions of
a reference and test image and then difference the images, we would not
expect to obtain a resulting image without systematic noise. What one would
invariably find from such a process is that at the position of each star
there would be systematic residuals. The degree and structure of these
residuals would be dependent upon the form of the spatial difference between
the PSFs. To achieve the best difference images one has to match the form of
the PSFs. As we mentioned earlier, the template is constructed so that it
has high S/N and seeing similar to the best observation\footnote{All observations
are well sampled with minimum FWHM of $\sim 2.8$ pixels.}. To match the
PSFs we degrade the reference images' seeing to match that of the test
observation.

The PSF-matching process is based on the fact that it is theoretically
possible to match the profiles of stars observed under two different sets of
conditions, with a simple convolution of the form given by equation
(\ref{simple}). Here $r$ characterises the flux distribution of a star in
good (better) seeing and $t$ that in poor seeing, $k$ is the convolution
kernel. A convolution in real space is equivalent to a multiplication in
Fourier space. Therefore, in principle, the Fourier Transform of the kernel
required to match the good seeing (reference) image to the poor seeing
(test) image, should be the quotient of the Fourier Transforms of the star.
The Inverse Fourier Transform recovers the required matching kernel
(equation \ref{ratio}).

\begin{equation}
t(x,y,z) = r(x,y,z) \ast k(x,y,z)
\label{simple}
\end{equation}

\begin{equation}
k = IFT \biggl( \frac{FT(t)}{FT(r)} \biggr)
\label{ratio}
\end{equation}

In reality a division of these Fourier Transforms is very sensitive to the
high frequency, low power, noise component. This leads to a poor match of
the images. To fix this one can use the fact that, the PSF of a star is
roughly Gaussian, and the Fourier Transformation of a Gaussian is a
Gaussian. We can then select a level in Fourier space below which the noise
component is dominant, and replace this with a Gaussian fit to the FT
(Ciardullo 1990). This method is not always useful because in many cases the
wings of the FT are not well modelled by a Gaussian.

To determine the best convolution kernel, the highest S/N, least
blended PSF stars are required. We thus need these {\it PSF stars} to be 
unblended with neighbouring stars in the bulge where effectively all 
stars are blended. Blended stars must be removed from those 
used in determining the PSF.

This situation was overcome by producing a list of a couple of hundred bright
stars on a given image. Stars were culled from the list based on the
relative proximity and brightness of neighbouring stars. Blending of
bright stars can in part be determined from the shape of the PSF of each
star. We thus selected stars based on their ellipticity, position angle,
FWHM and moment. The remaining stars were then combined to form a
generalised PSF profile for the image.

A further complication to the ideal case arises from the fact that the form
of the PSF varies across a frame because of poor focussing, telescope
flexure and temperature dependent effects (see \cite{tom96}). If a single
star is used to PSF-match an entire image, the systematic noise in the
difference image increases the further one gets from that star. A solution
to this is to split the image into sub-rasters and use a local PSF to match
the sub-regions of the image. The sub-rasters can then be mosaiced back
together to reform the image. Thankfully the magnitude of the PSF variation
for our observations is small, with modest sized effects due to differences only
being seen on the scale of $500$ $\times$ $500$ pixels (or $5'$ $\times$
$5'$).

The actual matching of PSFs is accomplished using the IRAF task {\em PSFmatch}. 
This task is similar to the implementation used by Riess et al. (1998) 
but has been updated to include new features. The convolution
kernel is determined for sub-rasters $500$ $\times$ $500$ and each is 
convolved separately then mosaiced to form 1k $\times$ 1k images. The entire
observation field of the MACHO camera is 4k $\times$ 4k per bandpass but
larger images were not made because of problems with matching such
as the scaling, colour terms, differences in bias levels and gain in the
different CCDs (\cite{rei98}).

The process of replacing the noise with the Gaussian fit is not regarded as
the optimal approach because of there are real differences between the PSF
shape and a Gaussian. To improve the difference image quality we
characterise the residuals in the difference image and remove them. This is
accomplished in the following manner. By stacking sub-rasters of the
difference image at the positions where the PSF stars were in the reference
image, we determined the median systematic residual at each pixel of the PSF
profile. This stacked residual image characterises the real deviations of
the PSF shape from that used in the matching process. This image was then 
scaled and subtracted from our test image PSF. We finally had an 
essentially empirical PSF and no need to replace the wings of the FT.
The convolution kernel is
now recalculated without the Gaussian replacement and the difference images
recalculated. This final PSF appeared to give the best possible subtraction with
the available stellar information. This step was found to reduce
the systematic residuals by a factor of $\sim\! 2$ in most cases. The final
resulting difference images had an average systematic noise component of
$\sim 1.3\%$ in both colours, with some images having less than $0.5\%$.

\subsection{Photometric Scaling}

A standard approach to matching reference and test images photometrically is
to make a linear fit to the pixel values for a single star found on both
images (\cite{rei98}). The test image is then scaled and offset 
using the fit values. We refer to this as a single point calibration.
Another approach is to use photometry taken on a matched set of stars in the
two images and scale the image based on this information.
In our analysis we apply this second approach which we believe is 
superior.

The single point technique has the advantage that it uses all the pixel
information for this region. However, it suffers from the fact that most of
the pixels have a low S/N ratio as they come from the wings or background
around a star. As there are very few pixels with high S/N, the fitted slope
can be skewed by the pixels near the background noise limit.

An important consideration when matching images is that the two
images were probably taken under different seeing conditions. Although
the PSF profiles have already been matched, real images can have gradients
in sky brightness and transmission. Sky brightness gradients can often be
attributed to the proximity of the observed field to the moon, and naturally
are dependent on its phase. Differences in the transmission come with
airmass, cloud cover and dust extinction. Such differences in sensitivity
can also be caused by problems with the flatfielding.

A single point calibration can not compensate for the presence of these
spatial variations. However, this can easily be achieved by performing
photometry on matched pairs of stars across the two images.  This photometry
gives a scaling factor (which represents changes in transmission), and an
offset term (which represent the change in the sky brightness), for each
position where the photometry was performed.  To deal with the possibility
of gradients we fit a low order polynomial\footnote{We expect variations in
sky level and transmission to be relatively smooth.} to the scaling and
offset terms and determine a transformation for each pixel in the test
image. Such a method requires a large number of points within the image to
constrain the fit. This is not a problem as there are many thousands of
stars per square arc minute in bulge fields.

The photometric scaling we used was always applied to the test 
image so that images were always registered to the same reference image.
In this way no correction was required for transmission differences with 
airmass.

\subsection{Object Detection}

The fully matched images were differenced to reveal objects whose flux rate
has varied in some way. The images were then searched for these sources.
Variable stars are detected as positive or negative sources as they became
brighter and fainter than they were in the reference image. However,
microlensing events and asteroids generally only appear as positive
excursions from the reference image baseline\footnote{It is possible for
microlensing events to appear with negative flux if they were
amplified in the combined reference images.}.

For source detection we applied a purpose written programme. We felt this
approach was necessary as the nature of the noise distribution is unlike
that in other images. The fundamental difference is that the noise varies
from pixel to pixel even though there is no signal. This is due to the
photon noise of the subtracted stars.  With such a programme we search for
positive and negative source simultaneously.

To detect objects we would ideally would like to know the noise at each
point in the image. To determine this we simultaneously examine the
reference and difference images. The significance of each detection is
accessed based on its S/N. We calculate the noise at each source position
using the flux in the reference image and systematics from the difference
image.  The various noise components (photon, readout, systematic) are well
characterised in the reduction and used to provide a position dependent
noise threshold for the difference image. The photon noise of the test
image dominates for faint objects whereas for brighter objects the
systematic noise from the difference image becomes important.  As the
seeing in our data set is always greater than two pixels, we choose to
determine the signal in a $3 \times 3$ pixel box at every point in the
image. The amount of signal in this box is compared with the noise expected
for the same pixels in the reference image.  Results with a S/N $> 3$
were written to a file along with other noise parameters such as the
systematic and readout noise of the difference image.

The detection process is carried out with images from
both passbands. The positions of objects found 
in the two colours are taken to be matched if they were
within $1''$. This colour match provides a strong 
constraint which separates the {\em real} objects from 
the large number of detections due to cosmic rays, bad 
pixels and saturation effects.

Candidate objects which pass these selection criteria were then checked
against those obtained in previous reductions of the field. If no match is
found, these new results were added to the database of object positions.  If a
result has previously been detected, the number of detections for this
object is incremented. This provides us with an extra parameter in
characterising the nature of the object. Objects detected most often can
usually be attributed to variable stars, whereas other detections
can be due to asteroids or microlensing events.

\subsection{Light Curves}

Aperture photometry is performed on the red and blue difference images at
the sources positions. We feel PSF photometry is unnecessary as the
difference images are no longer crowded with non-variable objects. The noise
is obtained from the initial images as noted in the detection strategy. With
this photometry we produce a database which is used to produce
difference-flux light curves for microlensing event detection.

These positions where photometry is performed are independent of whether
there is flux at this position in the reference image or not. This varies
from the traditional approach which has been to only follow up stars
detected in the template (reference) image. With this technique we are
searching among a database of variables for microlensing rather than among a
database of all objects.  The number of light curves which require scrutiny
is thus around a factor of twenty less than the number of stars detected in
the reference image.  The removal of the positional dependence gives us greater
sensitivity to detecting microlensing of stars which were either too faint
or too blended to be resolved in the reference image.  This effectively
provides us with a larger search area and a greater number of 
monitored stars.

\section{BLENDING}

Firstly, let us define what we mean by blending. This is the case in which
two or more stars are located within a few pixels of each other and the
seeing disk is greater than a few pixels. Such blending is dependent on
seeing and masks the true baseline flux of microlensing sources. In this
analysis, we are not concerned with the case where the lens in a
microlensing event emits flux which is blended with the flux from the
source. In this case the flux of the lens is not amplified and such blending
is generally negligible for stars with masses consistent with microlensing
results. The colour shifts associated with this type of blending usually
can not be detected with the present level of photometry (Buchalter et al. 
\cite{buc96}). This case is also generally inseparable
from blending with objects which are not the sources.  We are also
unconcerned with blending where the source is in a binary association. Here
the sources can be blended within the Einstein radius, but this constitutes a
small number of events(\cite{gri92}, \cite{dom98}) and in many cases one can
detect such events from the microlensing light curve.

The presence of the apparent blending of the stars monitored in microlensing
surveys has been known for some time (\cite{nem94}), and is inherent in the
the fact that the fields used for microlensing searches are crowded. These
fields were chosen because of the large number of stars that could be
monitored at one time. Crowding is demonstrated by figure \ref{crowd1},
where the number of stars blended within the seeing disk of the bright stars
can be clearly seen.

\placefigure{crowd1}

\notetoeditor{This figure should be two figures side by side in one column}

The main effect of the blending is that we do not know the true source flux
for most events. If such events are fitted as unblended sources, the
amplitudes of the events and the event time scales are underestimated, and
the number of stars monitored is under-estimated
(\cite{han97}, \cite{woz97}). The overall consequence, if blending is
unaccounted for, is that $\sim40\%$ of events towards the bulge are affected
by amplification bias and thus the determined optical depth can be
overestimated by a factor of $\approx 1.3 - 2.4$ (\cite{han97}). The
blending effect is therefore, a major factor in the determination of the true
optical depth to microlensing toward the galactic bulge, LMC, SMC and M31.
Statistical corrections have been made for the LMC and the bulge optical
depths (\cite{alc97,alc97b}), but corrections are really required on an
event by event basis. One way to bypass the effects of blending would be to
have HST images of all the events. With this approach we could resolve the
source star and obtain an unblended source flux. Unfortunately a large number
of hours of HST time are needed to perform this (\cite{han97a}). We note,
however, we do have HST observations of microlensing sources for a number of
important events.

To determine the blending in the case of classical microlensing, one must
attempt to fit the light curves to find the unlensed, blended, flux
component of the source baseline flux. In this regard Wo\'zniak \&
Paczy\'nski (1997) noted that one can only determine the blend fraction, with
the present accuracy of photometry, when the impact parameter is small,
$u_{min}$\(<\) 0.3. This is the case in only a small percentage of events.
The situation is not quite the same with DIA as we have different
information. We shall outline this in the next section. In short, instead of
the colour, centroid and light curve of the blend, we have the colour and
centroid of the source and the light curve of the source amplification.

\section{DIA VERSUS PSF PHOTOMETRY}

Our usual technique to detect microlensing is to perform PSF photometry
simultaneously at fixed positions in two passbands. With our DIA technique,
photometry is carried out at the end of the reduction process, at positions
where excess flux was detected at some time during the reduction.

The main advantage of using traditional PSF photometry over DIA comes when
dealing with fields where the stellar profiles are not blended together. In
such situations the determination of the PSF from a few stars is relatively
simple.  PSF photometry, unlike DIA photometry, is not subject to either the
addition of photon noise via the differencing process or the systematics
introduced in the image alignment and matching processes.

One common belief about difference image techniques, is that it is not as
powerful at constraining microlensing as standard PSF photometry because it
does not give a baseline flux. This is not the case. For 
classical microlensing events, photometry can simply be done once on the
reference image to determine the baseline flux.  In our present method of
analysis this image has a higher S/N than any individual image in the data
set, and has the best seeing and a well defined PSF. The photometry baseline determined from this frame can simply be
added to the individual difference frame photometry to provide the same flux
baseline as the standard PSF photometry. However, this source flux 
baseline can still be blended as in the PSF case.

To produce difference images with the lowest possible systematic noise, the
input images must be accurately photometrically matched. If the difference
image has only a small contribution from systematics, then we know that the
images were accurately matched photometrically. To perform PSF photometry we
have to make corrections for airmass and differences in transmission between
observations.

\placefigure{phot}

\placefigure{photb}

\notetoeditor{These figures should be on the same page.}

In figure \ref{phot} we show the photometry performed on an object with DIA
and with PSF photometry using the same set of images. A quite dramatic
improvement in the photometry is seen. Proof that this demonstrates a real
improvement in photometry is evident from the microlensing fit residuals
shown on the same figure. Although only one colour is shown, data points
from both passbands were consistent within the uncertainties presented. The
improvement in photometry comes partly from the fact that the source in this
event is highly blended with a neighbouring bright star. With DIA the nearby
non-variable star is more accurately removed than with the standard PSF
photometry. The PSF photometry is also dependent on how well the centroid
positions of the stars were initially determined in the template image. On
the other hand, DIA is dependent on how well the entire frame was aligned.
The flux from unlensed blended stars is removed accurately even when
unresolved with DIA. Such an improvement in the baselines of microlensing
events are required for the detection of the subtle effects of parallax and
the presence of binary sources.

Another advantage of the DIA scheme over PSF photometry is that the
difference flux represents the true colour of a microlensing event.  This
colour information is useful in deciding whether an event is blended or not.
Furthermore, all difference images are matched spatially and
photometrically. Thus we can stack images over the period where the
microlensed flux was above the background noise. The stacked difference
image can then be used to determine the source star colour more accurately.

The present solution to the problem of blending in the PSF photometry
is to use the fact that the measured flux within a microlensing 
event can be represented by equations \ref{Red1} and \ref{Blue1}
(\cite{alc97}),

\begin{equation}
f_{R}(t) = f_{uR} + A(t)f_{sR}
\label{Red1}
\end{equation}

\begin{equation}
f_{B}(t) = f_{uB} + A(t)f_{sB}
\label{Blue1}
\end{equation}

where $f_{uR}$ and $f_{uB}$ are the fluxes of the unlensed blended sources
and $f_{sR}$, $f_{sB}$ are the baseline fluxes of the lensed star.  All of
these terms must be found from the microlensing fits. The major problem with
this approach is that, even if we know the true colour of the source, we do
not know its true brightness. For bright, well covered microlensing events,
source fluxes can be found.  However, when uncertainties of $> 1\%$ are
associated with the photometry, the fit is practically degenerate with
respect to the unlensed flux component (see \cite{woz97}). This means we
can not accurately determine the value of the amplification for 
a large number of events.

In figure \ref{figx} we show three models of the amplitude of a microlensing
event with different values of the lensing parameters corresponding to
different values of source flux (given in table \ref{pars2}). In the top
panel there are three curves which are almost indistinguishable. In the
lower panel we plot the difference between the dashed curves and the solid
one. One can see that even for such large differences in source flux, hence
lensing amplitude, differences of less than a couple of percent occur in the
form of the difference in curve shape. The difference between the curves
increases only slowly as one moves away from the true values. 
This predicament is improved greatly with photometry taken in two or 
more colours because of the extra leverage this gives.

\placefigure{figx}
\placetable{pars2}

For DIA the flux in the two band-passes is given simply by equations 
\ref{Red2} and \ref{Blue2}. Here we do not have blending to 
consider so there is no unlensed flux term. The difference fluxes
in the two passbands are given by $f_{RD}(t)$ and $f_{BD}(t)$.
The constant $C$ is the ratio of blue to red flux which can be 
obtained from the colour of the source. 

\begin{equation}
f_{RD}(t) = (A(t)-1)f_{sR}
\label{Red2}
\end{equation}

\begin{equation}
f_{BD}(t) = (A(t)-1)Cf_{sR}
\label{Blue2}
\end{equation}

For this situation we only need to be able to measure the baseline flux
in one colour to determine the amplitude, since $Cf_{sR} \equiv f_{sB}$.  If
this is not possible, the accuracy of the determination of the amplitude
suffers from the same problems as the PSF photometry.  For cases where we
can not determine the baseline flux (pixel lensing), it should be possible
to estimate this quantity statistically.

\section{SOURCE FLUXES}

The case of determining the source flux with pixel microlensing is similar
to the case of determining the blending for classical microlensing.  In both
situations we do not know the initial source flux. The standard way to
determine the true amplitude of microlensing events is to use the shape of
the light curve. The accurate determination of the source flux requires well
sampled light curves with small percentage errors. In many cases the present
data does not meet this requirement.

With pixel lensing the source star of the event is initially unresolved and
the S/N is generally low. The task of determining an accurate source flux
can be impossible in many cases. However, from the reference image we can
determine an upper limit to this flux. In such cases, as for DIA of
classical microlensing events, we still have an accurate position and colour
for the source. With this colour (and the associated uncertainties), the HST
luminosity function of the bulge (\cite{hol98}), the upper limit of source
flux, and the shape of the light curve, one can determine the probability
distribution associated with the source flux for each pixel lensing event.
This source flux probability distribution and the light curve shape can be
used to give the distribution of $\hat t$ for each event. The combination
of these $\hat t$ distributions gives us an overall distribution of
$\hat t$ from which the lens mass distribution can be extracted in the
traditional way (Griest 1991).

For classical microlensing events where the sources are faint and the
amplifications are moderate, the associated photometry errors are typically
greater than a few percent. Typically for these events, as for pixel lensing
events, we can not accurately determine the amount of unblended lensed light
because of blending. In order to use these events when determining the
microlensing optical depth, one approach could be to use the same technique
as outlined above for pixel lensing. This would help to constrain a large 
number of classical events. Indeed this approach is desirable when we 
consider the effect of blending on measured optical depth.

Aside from the aspects mentioned above, we note that, Griest and Hu (1992) and
Dominik (1998) found that it was possible that galactic binary sources are
fitted well with a single source with blending. As DIA does not suffer from
blending this caveat is removed and the light curves should thus readily
give unbiased results for binary sources. 

Again we remind the reader to the results of this analysis can be
found in paper 2.

\section{SUMMARY}

We have outlined a detailed approach to the DIA scheme.  A method of using
multiple PSFs to addresses the shortfalls of matching images with a single
PSF was discussed. We noted how gradients in transmission and sky level
across an image can be important when producing difference images. A
solution using a spatially dependent photometric offsets and scaling was
outlined. A method for reducing difference images noise by combining
observations to form the reference image was presented.

The importance of using an accurate image registration was stressed.
The effect of differential refraction on difference images
has been examined and is shown to be crucial.  A method for compensating for
differential refraction by offsetting stars relative to their colours was
presented.

A new method for determining the distribution of event time scales
was discussed. This method would use existing HST colour-magnitude
diagrams to determine the distribution of possible event sources 
with colour. The colours of events determined with DIA would then
be used to determine a distribution of possible $\hat t$ values
for each event. These would then be combined to determine the overall
distribution of $\hat t$ values. From this lens mass distribution 
can be found in the usual way.

We have demonstrated how the DIA photometry can make a large improvement
in the quality of light curves over PSF photometry because of blending.
We discussed how there are fundamental differences between results from 
PSF and DIA photometry. Difference images naturally provide unblended 
colours and centroid positions and lightcurves for microlensing events, 
while PSF results are usually blended to some extent.

We are grateful for the skilled support by the technical staff at Mount
Stromlo Observatory. Work at Lawrence Livermore National Laboratory is
supported by DOE contract W7405-ENG-48. Work at the center for Particle
Astrophysics at the University of California, Berkeley is supported by NSF
grants AST 88-09616 and AST 91-20005. Work at Mount Stromlo and Siding
Spring Observatories is supported by the Australian Department of
Industry, Technology and Regional Development. Work at Ohio State
University is supported in part by grant AST 94-20746 from the NSF. 
W. J. S. is supported by a PPARC Advanced Fellowship. K. G. is grateful 
for support from DOE, Sloan, and Cottrell awards. C. W. S. is grateful 
for support from the Sloan, Packard and Seaver Foundations.
This work was carried out by A.J.D. in partial fulfilment of the 
requirements for the degree of PhD at ANU.


\newpage
\pagestyle{empty}

\newpage

\newpage

\newpage

\newpage
\begin{figure}
\epsscale{2.2}
\caption{The effect of differential refraction on
difference images. Left image is a difference image without applying any
correction for differential refraction effects. The right image is the
same image with our correction technique applied. The scale of the noise
structures is much reduced although not completely removed. The two images
are $100'' \times 100 ''$. The residual object in the centre is due to a
variable star\label{figdiffl}}
\end{figure}

\newpage
\begin{figure}
\epsscale{1.0}
\caption{Blending. Left: the position of an
microlensing event (arrowed) in an $11'' \times 11''$ ground based image
Macho database image.  Right: a drizzled HST image of the same region with
reference stars (A - E) to approximately the same scale. Note the number
of objects blended within the ground based images seeing disk. This result
comes from an event in the LMC where the crowding is similar to the images
analysed in this analysis.\label{crowd1}}
\end{figure}

\newpage
\begin{figure}
\epsscale{1.0}
\plotone{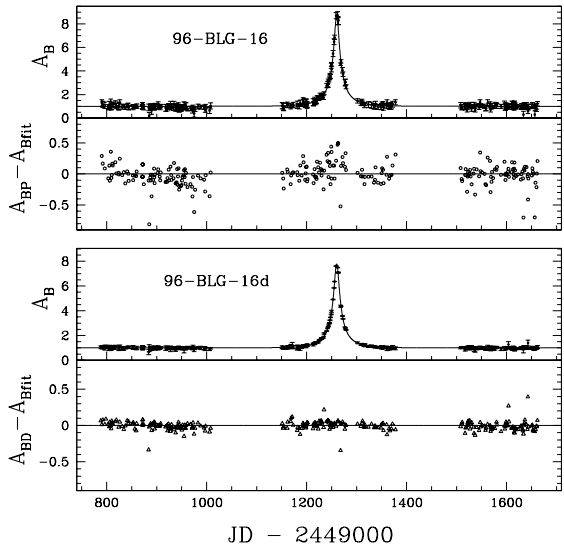}
\caption{This figure shows the light curves for microlensing
event 96-BLG-16. The top two panels are the standard macho 
blue ($B_{m}$) and red ($R_{m}$) PSF photometry and microlensing
fit residuals. The bottom two panels are the DIA photometry 
and residuals for the same initial data. The fits vary slightly 
because the PSF photometry is blended and the difference image 
photometry is not. Further results are given in paper 2 (\cite{alc99}).
\label{phot}}
\end{figure}

\newpage
\begin{figure}
\epsscale{0.95}
\plotone{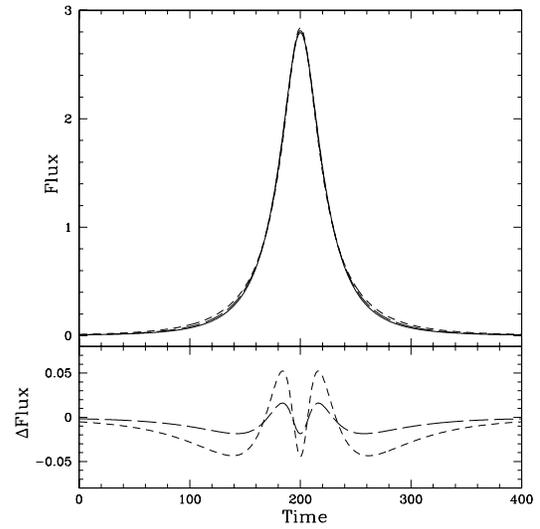}
\caption{Differences between microlensing light curves for a range of
source fluxes. In the top panel we show three theoretical microlensing 
light curves for three values of source flux. In the lower panel we 
show the difference between the dashed curves and the solid curve.
We can readily see that, to determine the differences between these 
light curves and hence true source flux, we require very small 
uncertainties and good sampling. The parameters are given in 
table~\ref{pars2}.\label{figx}}
\end{figure}

\clearpage
\begin{deluxetable}{crrrrr}
\tablewidth{0pt}
\tablecaption{Parameters of Model Events.\label{pars2}}
\tablehead{
\colhead{$f_{s}$}& \colhead{$A$}  & \colhead{$U_{min}$} & 
\colhead{$\hat{t}$} & \colhead{$t_{max}$}
}
\startdata
2.0 & 2.40 & 0.448 & 85.4 & 200\nl
1.5 & 2.87 & 0.365 & 98.0 & 200\nl
1.0 & 3.83 & 0.268 & 120.8 & 200\nl

\enddata
\tablecomments{Microlensing parameters for simulated events presented 
in figure ~\ref{figx}.}
\end{deluxetable}

\end{document}